\documentclass[aps,prb,twocolumn,amsmath,amssymb,gallery,footinbib,superscriptaddress,floatfix]{revtex4-2}

\usepackage{graphicx}
\usepackage{dcolumn}
\usepackage{bm}
\usepackage{hyperref}

\begin{document}

\preprint{APS/123-QED}

\title{Kittel law and domain formation mechanism in PbTiO$_{3}$/SrTiO$_{3}$ superlattices.}

\author{Fernando Gómez-Ortiz}
\affiliation{Departamento de Ciencias de la Tierra y Física de la Materia Condensada, Universidad de Cantabria, Avenida de los Castros s/n 39005 Santander. Spain.}
\author{Hugo Aramberri}
\affiliation{Materials Research and Technology Department, Luxembourg Institute of Science and Technology (LIST), Avenue des Hauts-Fourneaux 5, L-4362 Esch/Alzette, Luxembourg}
\author{Juan M. L\'opez}
\affiliation{Instituto de F\'{\i}sica de Cantabria (IFCA), Universidad de Cantabria-CSIC, 39005 Santander, Spain}%
\author{Pablo García-Fernández}
\affiliation{Departamento de Ciencias de la Tierra y Física de la Materia Condensada, Universidad de Cantabria, Avenida de los Castros s/n 39005 Santander. Spain.}
\author{Jorge Íñiguez}
\affiliation{Materials Research and Technology Department, Luxembourg Institute of Science and Technology (LIST), Avenue des Hauts-Fourneaux 5, L-4362 Esch/Alzette, Luxembourg}
\affiliation{Department of Physics and Materials Science, University of Luxembourg, Rue du Brill 41, L-4422 Belvaux, Luxembourg}
\author{Javier Junquera}%
\email[Corresponding Author:~]{javier.junquera@unican.es}
\affiliation{Departamento de Ciencias de la Tierra y Física de la Materia Condensada, Universidad de Cantabria, Avenida de los Castros s/n 39005 Santander. Spain.}

\date{\today}

\begin{abstract}
We report second-principles simulations on the structural and energetic properties of domains in (PbTiO$_{3}$)$_{n}$/(SrTiO$_{3}$)$_{n}$ superlattices.
For the explored layer thickness ($n$ ranging between 8 and 16 unit cells) and lateral sizes of the domains, the most stable configuration corresponds to polar domains separated  by  a  sequence  of  counter-rotating  vortices  (clockwise/counterclockwise) perpendicular to the stacking direction and acting as domain walls.
The balance between the domain wall energy and the electrostatic energy yields to an optimal domain period $\omega$ that is proportional to the square-root of the thickness of the PbTiO$_{3}$ layer, following the Kittel law.
For a given lateral size of the simulation box, suboptimal domain structures (with a width larger than the one predicted by the Kittel law) can be obtained in a metastable form.
However, at finite temperature, molecular dynamics simulations show the spontaneous change of periodicity, which implies the formation of new domains whose generation is initiated by the nucleation of vortices and antivortices at the interface between the SrTiO$_{3}$ and the PbTiO$_{3}$ layers.
The vortices progressively elongate and eventually annihilate with the antivortices yielding the formation of new domains to comply the Kittel law via a topological phase transition.
\end{abstract}
\maketitle

\section{Introduction}
\label{sec:introduction}

A common feature among the family of ferroelectric materials is the formation of domain structures, regions of space with different polarization separated by boundaries called domain walls~\cite{Tagantsev-book,Catalan12}. 
Domains of opposite polarization lead to an overall charge neutrality at the surfaces reducing the depolarization field and the associated electrostatic energy. 

The structure and energetics of domains in ferroic materials were first addressed by Landau and Lifshitz~\cite{Landau-35}, and one decade later by Kittel~\cite{kittel1946,Kittel-49} in his studies on ferromagnetic domains.
The delicate balance between the energy of the boundary between domains, the magnetic field energy of the configuration, and the anisotropy energy of the spin orientation determine the relationship between the width of the domains, $\omega$, and the thickness of the material, $d$~\cite{kittel1946}. %
Adding up all the energy costs and minimizing this with respect to the domain size leads to a square-root dependence of $\omega$ as a function of $d$. This is the so-called Landau-Kittel law,  $\frac{w^2}{\delta}=A\cdot d$, where $A$ is an adimensional proportionality constant and $\delta$ is the thickness of the domain wall. 
This law was extended to ferroelectric materials by Mitsui and Furuichi~\cite{Mitsui-53} studying the domain structure of the Rochelle salt, where the electrostatic, elastic, and gradient energies determine the number and width of the domains for a given thickness of the material.
The square-root dependence was further generalised under specific periodicities and screening conditions to the case of ultrathin ferroelectric layers~\cite{Catalan-08} and to the case of superlattices with paraelectric materials~\cite{Bennett-20}.
Moreover, Roitburd expanded it also for ferroelastic thin films under epitaxial strain~\cite{Roitburd-76}. 
Therefore, it seems that the Landau-Kittel law of stripe domain width on film thickness is a general property of all ferroics~\cite{Catalan12}.
\vspace{-0.9mm}

Beyond the analytical derivations, the validity of the law has been confirmed by first-principles-based studies in ferroelectric~\cite{Lai-07.3} and multiferroic~\cite{Prosandeev10} thin films with thicknesses down to three unit cells. 
Following the spirit of these two works, we widen the applicability of the Kittel law to the case of ferroelectric/dielectric superlattices characterized by a ground state consisting of polar domains separated by a sequence of counter-rotating vortices (clockwise/counterclockwise) acting as domain walls~\cite{Aguado-Puente-12,Yadav-16,Shafer-18}.
Using second-principles simulations we validate the law for this complex polarization texture. 
Very interestingly we show how when the system is initialized from a metastable state, where the density of domains is smaller than the one predicted by the Kittel law, it evolves upon heating to the ground state via a topological phase transition.
The driving mechanism for the generation and closure of new domains is the recombination of vortex and antivortex defects generated at the interface between PbTiO$_3$ and SrTiO$_3$.

\section{Methodology}
\label{sec:methodology}
The  second-principles  simulations  were  performed  using  the  same  methodology  presented  in  previous works~\cite{Wojdel2013,Zubko-16}, as implemented in the {\sc{scale-up}} package~\cite{Wojdel2013,Pablo2016}. 
The second-principles parameters of both materials were fitted from density
functional theory imposing a hydrostatic pressure of $-11.2$ GPa to counter the
underestimation obtained by the local density approximation of the cubic lattice constant
that was taken as the reference structure. We imposed an epitaxial constraint
assuming in-plane lattice constants of $a=b= 3.901$~\AA, forming an angle $\gamma= 90^{\circ}$ mimicking the conditions of a SrTiO$_3$ substrate.
The interatomic potentials, and the  approach  to  simulate  the  interface,  are  the  ones first introduced in Ref.~\cite{Zubko-16}. 
For a given value of supercell periodicity $n$, (PbTiO$_3$)$_n$/(SrTiO$_3$)$_n$ several values of lateral size $L$ were relaxed making them commensurate with the number of simulated domains. We solved the models by running Monte Carlo-simulated annealing from $60$~K down to very low temperatures, typically comprising 20,000 relaxation sweeps.
Regular Langevin molecular dynamics methods at constant temperature of $T=90$~K were also used to solve the models in order to follow the dynamics of the emergent domains. 
For computational feasibility, we have focused on a simulation supercell made of a periodic repetition of $L\times1\times2n$ elemental perovskite unit cells for the Monte Carlo-simulated annealings and of $L\times10\times2n$ for following the dynamics of the domains. As proven in~\cite{Gomez-Ortiz22}, at low temperatures ($T<73$~K), the vortices do not vary along the axial $y$-direction. Therefore, the simplification of taking one unit cell along this direction does not affect the validity of the model while it speeds up the calculations. However, when $T=90$~K a sufficiently high number of unit cells along the $y$-direction must be considered in order to account for the variation along the axial direction. 

We obtained the force-constant band calculations using the direct supercell approach as implemented in the {\textsc{phonopy}} package~\cite{Togo-15}. To this end, we considered the high-symmetry unit cell of the superlattices (in which the atoms in the PbTiO$_{3}$ and SrTiO$_{3}$ layers occupy the cubic-like perovskite positions), and repeated it 4$\times$4 times in the $xy$-plane to build the supercell for the calculations, which we found to be large enough to yield well-converged results. We included the non-analytical contribution to the bands which accounts for the splitting between the longitudinal and transverse polar bands as implemented in~\cite{Gonze-97}.

Local polarizations are computed within a linear approximation of the product of the Born effective charge tensor times the atomic displacements from the reference structure positions divided by the volume of the unit cell.

\section{Results}
\label{sec:results}

\subsection{Validation of Kittel law}
\label{sec:validKittel}

\begin{center}
  \begin{figure*}[!]
     \centering
      \includegraphics[width=\textwidth]{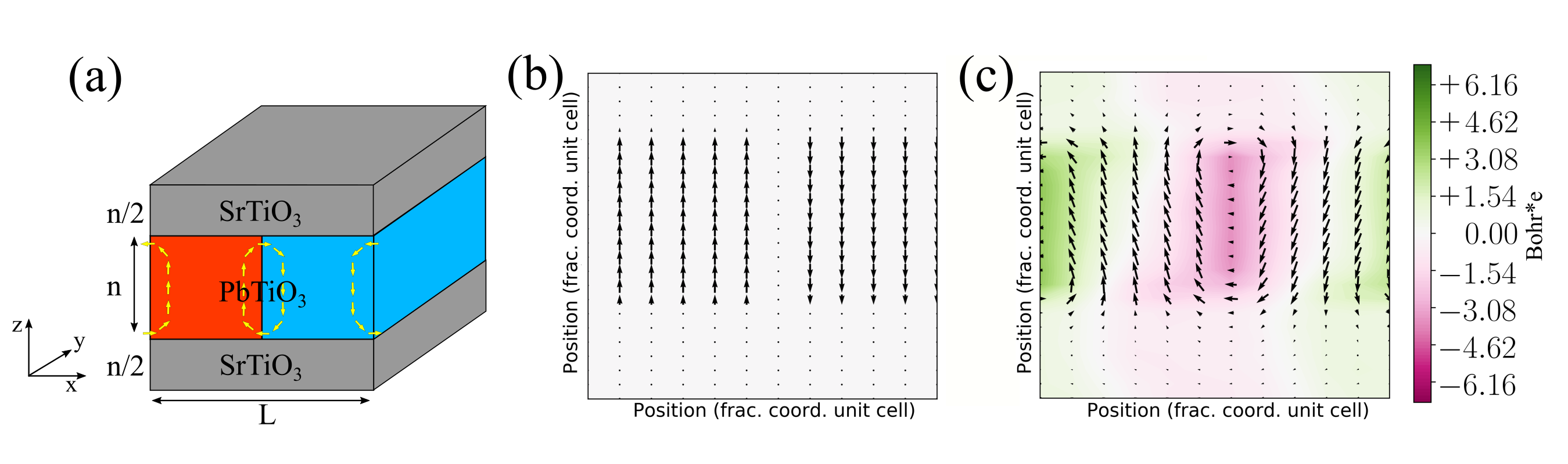}
      \caption{Structural relaxation of a two-domain structure in (PbTiO$_3$)$_n$/(SrTiO$_3$)$_n$ superlattices. (a) Schematic view of the simulation cell periodically repeated in the three directions of the space. Red (blue) regions indicate the positive (negative) polarization domains along the $z$ direction. (b) Local dipoles of the initial Ising-like configuration presenting a clear two-domain structure for a layer thickness of $n$ = 12, and a lateral size $L=12$ u.c. (c) Corresponding pattern of local dipoles after relaxation, showing an alternating clockwise/counterclockwise polar vortex configuration~\cite{Shafer-18} along $x$. The domain wall thickness $\delta$ extends over one unit cell. Colors represent the axial component of the polarization, perpendicular to the plane defined by the vortices.}
      \label{fig:Figure-1} 
  \end{figure*}
\end{center}

\begin{center}
  \begin{figure}[b!]
     \centering
      \includegraphics[width=\columnwidth]{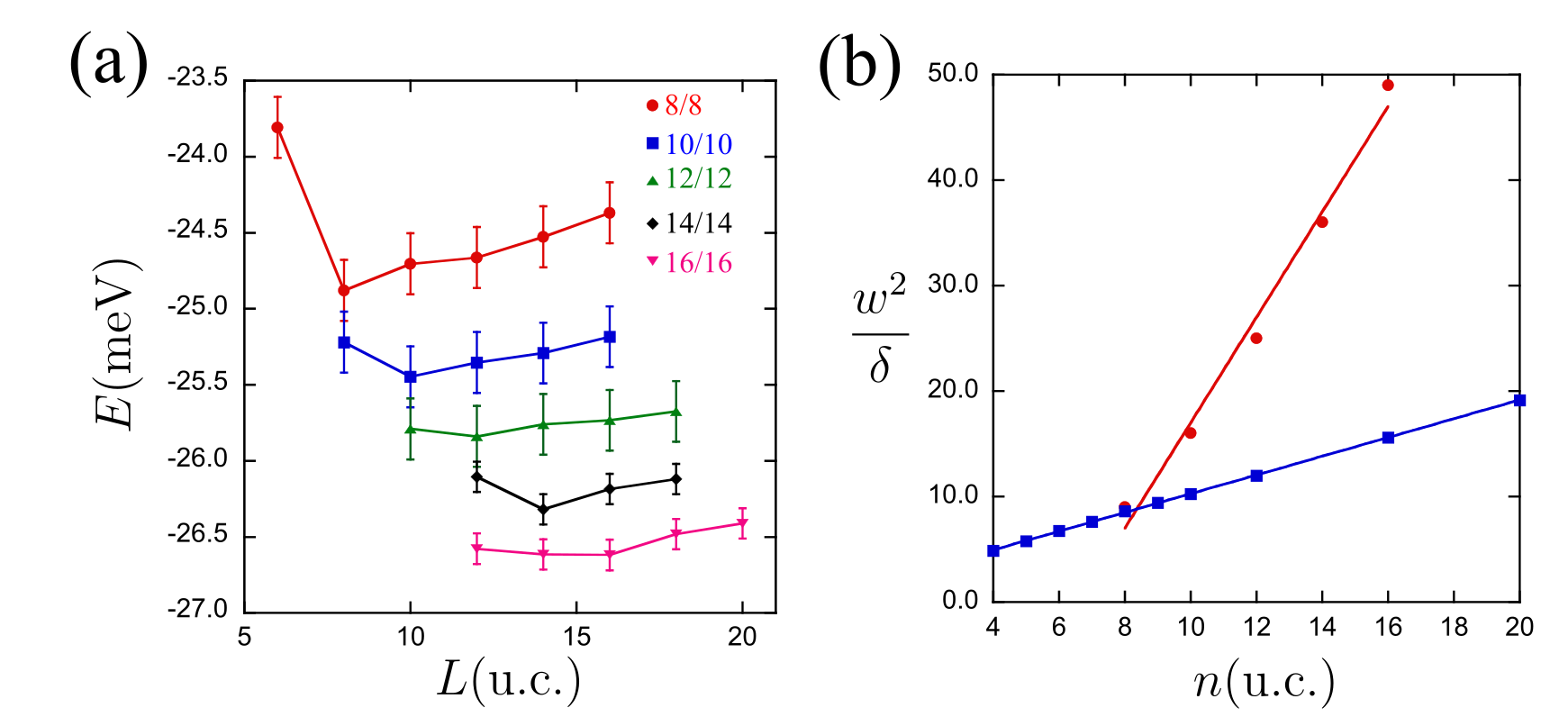}
      \caption{Optimization of the lateral size of a two-domain structure in (PbTiO$_3$)$_n$/(SrTiO$_3$)$_n$ superlattices. (a) Energy per five atoms unit cell as a function of the lateral size of the supercell for different periodicities: red dots ($n=8$), blue squares  ($n=10$), green up-triangles  ($n=12$), black diamonds  ($n=14$), and magenta down-triangles  ($n=14$). (b) Linear fit of the squared optimized width of the domain as a function of the layer thickness by two procedures. Red dots correspond to minimizing the total energy of the supercell while blue squares correspond to the minimum of the force-constant bands along $\Gamma - X$.}
      \label{fig:Figure-2} 
  \end{figure}
\end{center}

 We have checked the validity of the Kittel law in our superlattices following a similar recipe as in Refs.~\cite{Lai-07.3,Prosandeev10},
For different layer thicknesses $n$ with constant and equal dielectric/ferroelectric ratio, ranging between 8 and 16 unit cells, the lateral size of the supercell to host two domains was optimized [see Fig.~\ref{fig:Figure-1}(a)].
In order to achieve this goal for every value of $n$ different lengths of the supercell along the $x$-direction, $L$, were simulated. 
Once $n$ and $L$ were fixed, the initial atomic positions were chosen to mimic a couple of pure Ising domains, where the polarization changes abruptly from pointing upwards  (up-domain) to downwards (down-domain) along the $z$-direction in just one unit cell, as shown in Fig.~\ref{fig:Figure-1}(b). 
This configuration was taken as the starting point of the Monte Carlo annealing.
The resulting typical dipole configuration, a local minimum at low temperature, is shown in Fig.~\ref{fig:Figure-1}(c).
The spontaneous formation of alternating pairs of clockwise/counterclockwise vortices along the $x$-direction is clearly visible, together with the development of an axial component of the polarization along the $y$-direction.
These vortices were already theoretically predicted from phenomenological theories~\cite{Stephenson-06, Bratkovsky-09}, 
first-principles-based effective Hamiltonian~\cite{Kornev-04}, second-principles simulations~\cite{Zubko-16,Shafer-18,Gomez-Ortiz22} or full first-principles calculations~\cite{Aguado-Puente-12}, and experimentally demonstrated~\cite{Yadav-16} in PbTiO$_{3}$/SrTiO$_{3}$ superlattices.
For a given layer thickness, the energy per five-atom unit cell as a function of the lateral size of the supercell, always assuming the presence of two domains in the simulation box, is shown in Fig.~\ref{fig:Figure-2}(a). 
The first observation that can be drawn is that the larger the layer thickness, $n$, the smaller the energy per unit cell. 
This fact stems from two different causes.
On the one hand, the polarization in the PbTiO$_{3}$ layers increases with $n$, tending to the bulk polarization value and approaching the ground state of the domain.
On the other hand, the larger the layer thickness the smaller the polarization within the SrTiO$_{3}$ layer; the system undergoes a transition from an electrostatically ``coupled'' regime for small $n$ to a ``decoupled'' regime for large $n$~\cite{Stephanovich-03,Stephanovich-05,Zubko-12}. Since the SrTiO$_{3}$ layers will be closer to the ground-state unpolarized configuration, the energy is also reduced.
The second observation that can be drawn is that the energy curve of the two-domain structure as a function of the lateral size $L$ presents a minimum corresponding to the most stable geometry. The larger the layer thickness the shallower the minimum, that is localized for longer values of $L$.
From these minima, we can infer the optimal width of the domain, $\omega$. Assuming that the domain wall is one-unit-cell thick [in good agreement with our simulations; see Fig.~\ref{fig:Figure-1}(c) for a typical case], $\omega = L/2 -1$ unit cells. 
Plotting the square of these optimal widths against the thickness of the PbTiO$_3$ layer, red dots in Fig.~\ref{fig:Figure-2}(b), we recover a linear behavior as predicted by the Kittel law, a tendency also shown in BiFeO$_{3}$~\cite{Prosandeev10} and Pb$($Zr,Ti$)$O$_3$ ultrathin films~\cite{Lai-07.3}.

\begin{center}
  \begin{figure*}[!]
     \centering
      \includegraphics[width=13cm]{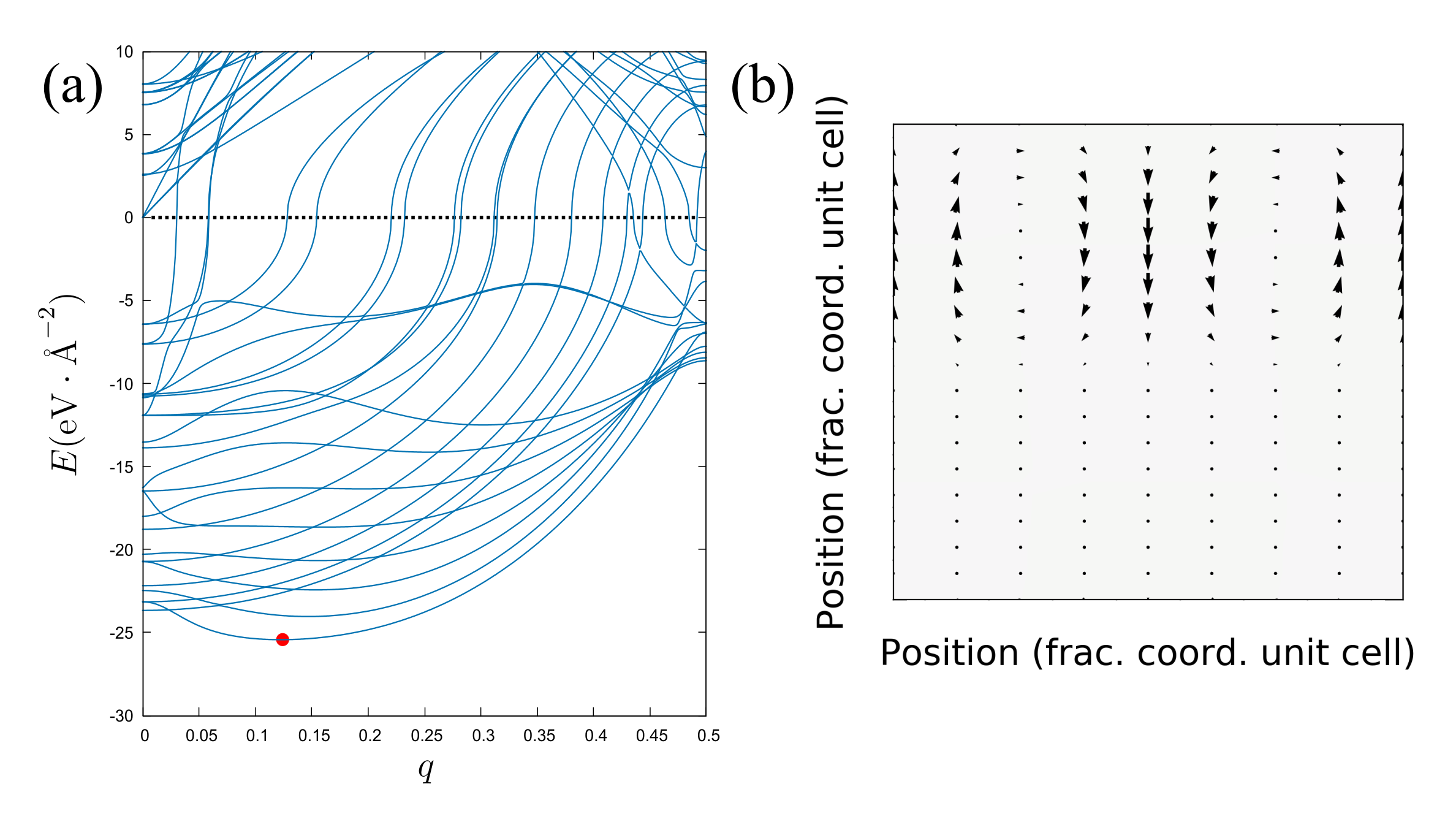}
      \caption{(a) Force-constant bands along $\Gamma -X$ obtained after diagonalization of the force-constant matrix in a\linebreak (PbTiO$_3$)$_9$/(SrTiO$_3$)$_9$ supercell. Dashed line indicates the energy of the centrosymetric configuration which is taken as the reference. Red dot indicates the position of the strongest instability located at $q_{\rm min}=0.123$ in fractional units.(b) Relaxed structure following the strongest instability of the force-constant bands.}
      \label{fig:Figure-3} 
  \end{figure*}
\end{center}

Alternatively, one can try to predict the optimal width of the multidomain structure by studying force-constant bands like the ones presented in Fig.~\ref{fig:Figure-3}(a). Following the unstable modes along $\Gamma -X$ we can identify $q_{\rm min}$ as the wave-vector associated with the strongest instability describing vortex structures as the ones presented in Fig.~\ref{fig:Figure-3}(b). From this value we can infer the optimal width as $\omega=\frac{1}{2q}-1$ [see blue squares in Fig.~\ref{fig:Figure-2}(b)]. 
Interestingly, while the square-root dependence of the domain width as a function of the layer periodicity is nicely reproduced, there exists a discrepancy with the results obtained via the first method. The (harmonic) force-constant analysis predicts narrower domains than the ones obtained from a full energy minimization.
This difference must be due to anharmonic effects and/or the fact that the fully relaxed structures feature a combination of phonon mode distortions to optimize the energy. Thus, for example, the development of a slight offset coupled with an in-plane component of the polarization as the one shown in Fig.~\ref{fig:Figure-1} (c) and experimentally attained \cite{Yadav-16}, although not captured in the force-constant analysis [see how vortices are centered in Fig.~\ref{fig:Figure-3}(b)], reduces the normal component of the polarization to the surface. The tilt of the polarization reduces the depolarization charges at the surface and allows the widening of the domains, resulting in the underestimate of the domain width predicted by the harmonic analysis.
\subsection{Domain formation}
\label{sec:domainformation}
\begin{center}
  \begin{figure}[h]
     \centering
      \includegraphics[width=\columnwidth]{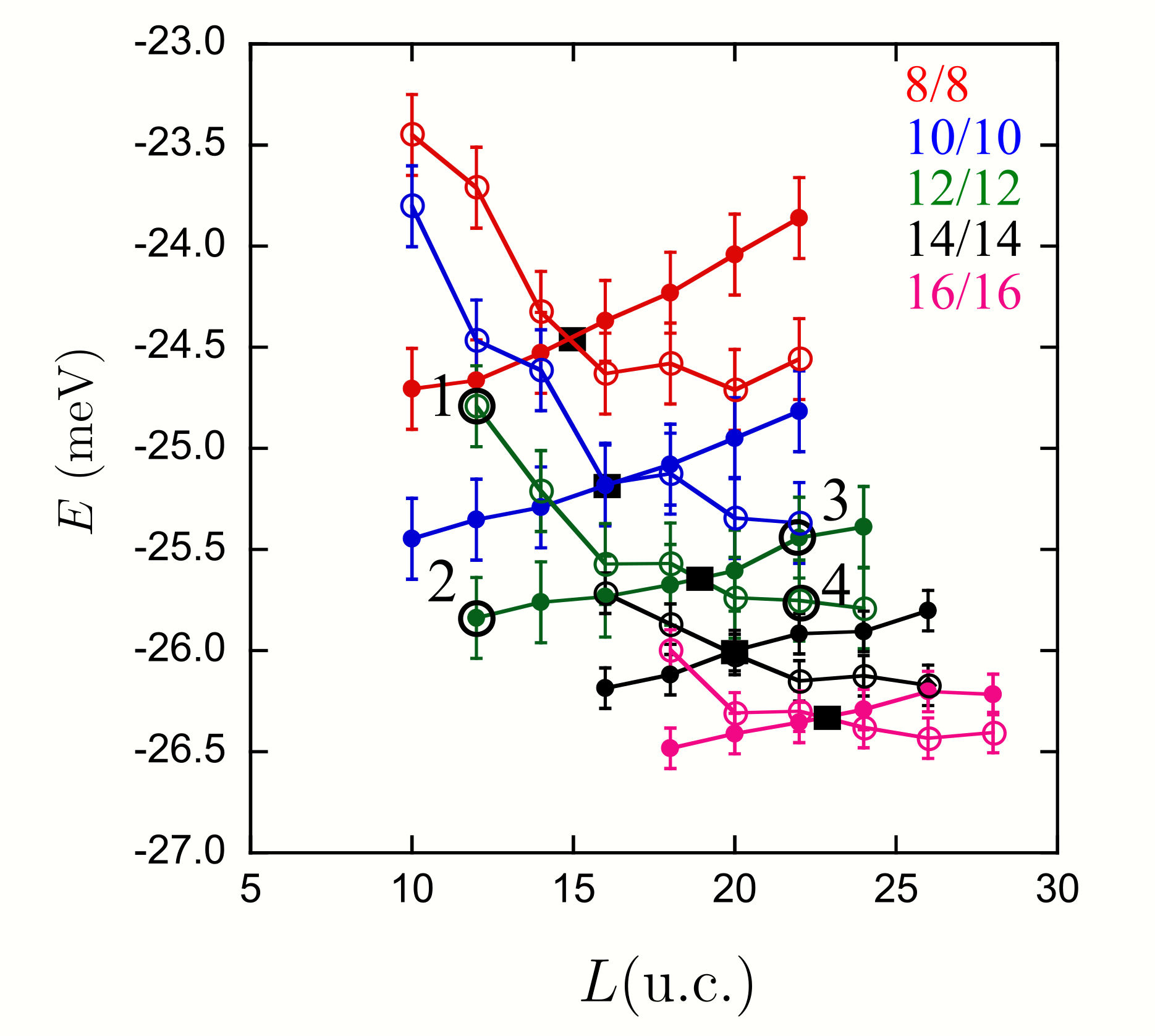}
      \caption{Energy profile per five atom unit cell as a function of the lateral size of the supercell for the two- (filled dots) and four- (open dots) domain structures. Different layer thicknesses are indicated by colors: red ($n =8$), blue ($n =10$), green ($n =12$), black ($n =14$) and magenta ($n =16$). Filled black squares indicate the crossing point where the four-domain structure becomes more stable. Numbers 1-4 label the different dipole patterns plotted in Fig.~\ref{fig:Figure-5}.}
      \label{fig:Figure-4} 
  \end{figure}
\end{center}

\begin{center}
  \begin{figure}[h]
     \centering
      \includegraphics[width=\columnwidth]{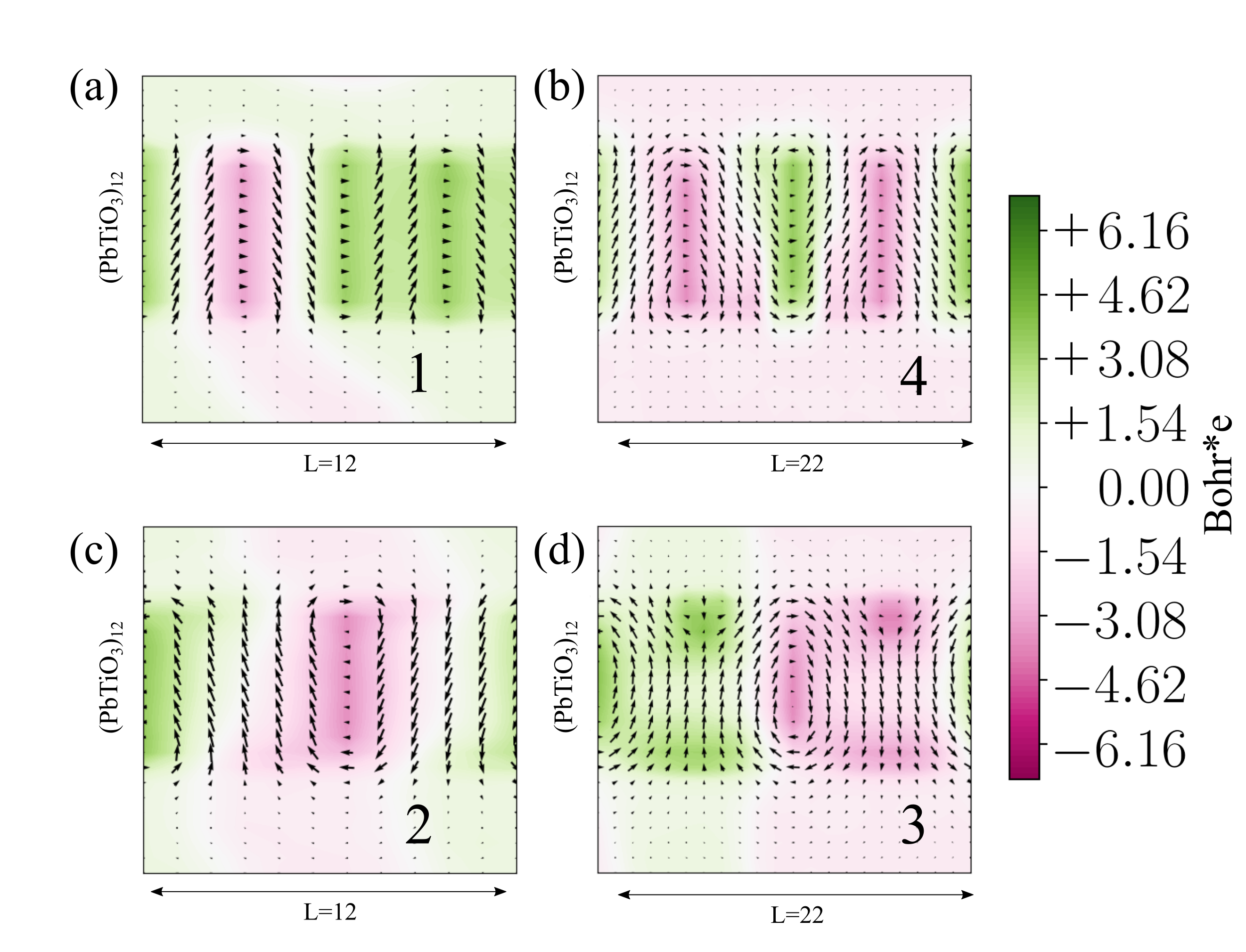}
      \caption{Polarization map of the relaxed structures labeled from 1 to 4 in Fig.~\ref{fig:Figure-4} for different lateral sizes $L$ and a layer thickness of $n=12$. Black arrows indicate the local dipoles, projected onto the $(x,y)$ plane. The axial component of the polarization along the $[010]_{\rm pc}$ direction is represented by the green and magenta color map. Units of the dipoles in  $e \times$ Bohr, where $e$ is the electron charge.}
      \label{fig:Figure-5} 
  \end{figure}
\end{center}

Up to now, the existence of two domains in the simulation box has been imposed in the calculations. 
In the following we shall study the phase competition between different configurations under suboptimal domain widths. 
This is shown in Fig.~\ref{fig:Figure-4} where we compare the energy per unit cell considering two or four-domains for a given layer thickness, $n$, and lateral size of the supercell, $L$.
Again, for each pair of $(n,L)$ a Monte-Carlo annealing was carried out starting from  purely 180$^\circ$ Ising-like structures containing two- or four-domain walls in the simulation box.
In every case, the final relaxed structures display the typical polar vortex configuration, similar to the one shown in Fig.~\ref{fig:Figure-1}(c). 
In Fig.~\ref{fig:Figure-4} we plot the evolution of the energy profile per five-atoms unit cell as a function of the lateral size of the supercell. 
As already discussed in Fig.~\ref{fig:Figure-2}(a), for a given amount of domain walls in the supercell, the larger the layer thickness $n$ the smaller the total energy.
But the most important conclusion that can be drawn from Fig.~\ref{fig:Figure-4} is that, for a given $n$, a crossover between the two- and four- domain configurations is found for a critical length $L_{\rm c}$, whose value increases with $n$, as marked by the filled squares in Fig.~\ref{fig:Figure-4}. 
Studying lateral sizes far from the coexistence regions (below or above $L_{\rm c}$) different characteristic features can be observed for the evolution of the unstable phases.
In Fig.~\ref{fig:Figure-5} we show the behaviour for a layer thickness of $n = 12$.
Below $L_{\rm c}$ (short lateral sizes) the four-domain structure is only metastable since the large penalty coming from the gradient energy term is not compensated by the saving in electrostatic energy.
Indeed, in this metastable regime the system tends to reduce the energy gradient contribution by forming polarization waves~\cite{Lu-18} [Fig.~\ref{fig:Figure-5}(a)] along the $[100]_{\rm pc}$ direction, with the concomitant onset of a net in-plane polarization, and the displacement of the center of the vortices in the PbTiO$_{3}$ layer towards the interface with SrTiO$_{3}$. The smaller the lateral size of the supercell $L$, the larger the offset between neighboring vortex cores that do not fit in the center of the PbTiO$_3$ layer as in the case of larger lateral sizes [see Fig.~\ref{fig:Figure-5}(d)]. 
This phenomenon has also been observed in BiFeO$_3$ ultrathin films~\cite{Prosandeev10}. 

\begin{center}
  \begin{figure}[h]
     \centering
      \includegraphics[width=\columnwidth]{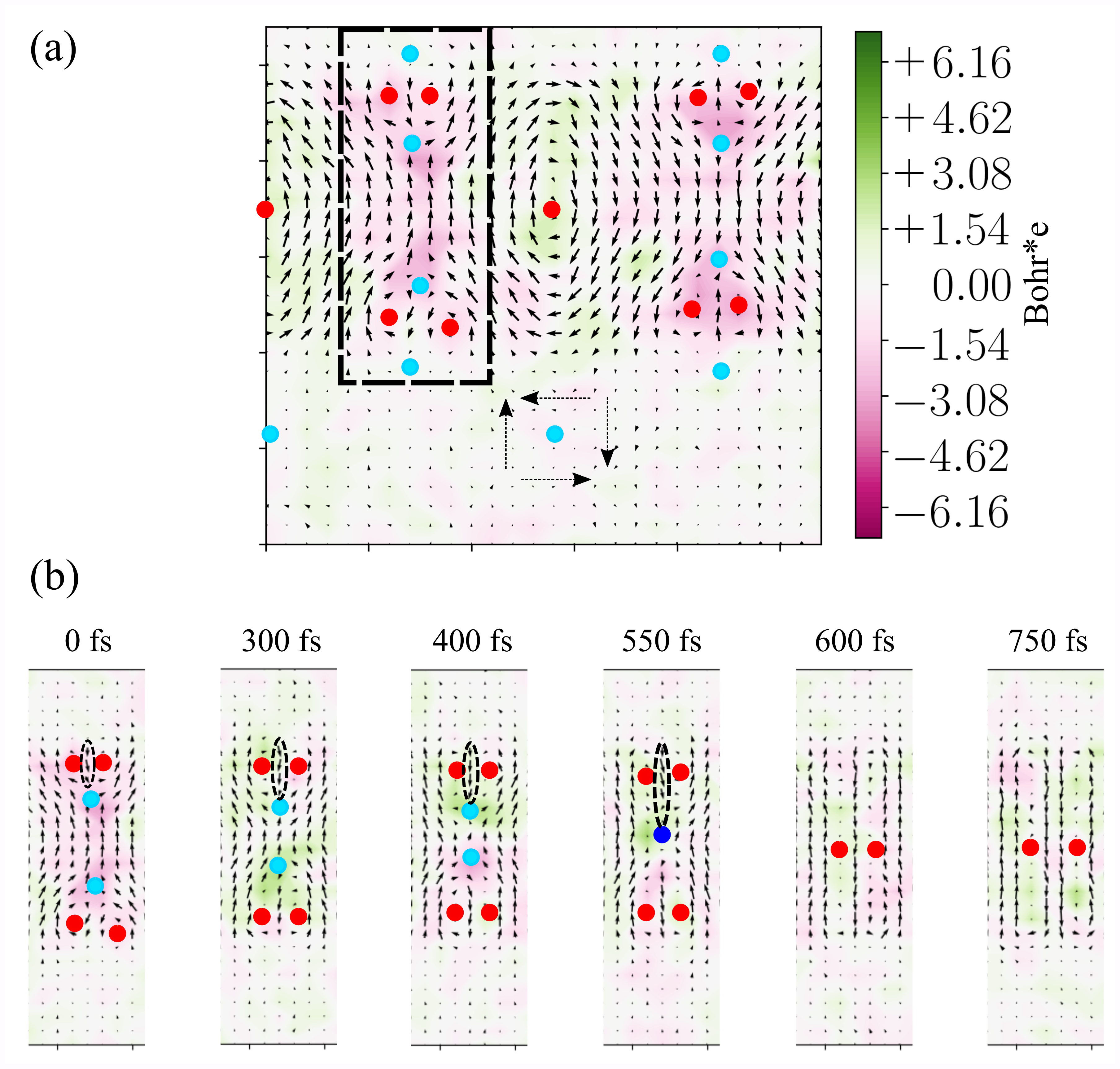}
      \caption{Polarization map for a $n=14$, $L=28$ supercell at finite temperature ($T = 90$ K) and compressive epitaxial strain of $-0.5$\%. (a) Initial two-domain structure configuration. Dashed square delimits the upwards domain studied for the dynamics in (b), dashed arrows within the SrTiO$_3$ are a guide to the eye to locate one of the different antivortex structures. 
      (b) Temporal evolution obtained by molecular dynamics simulations at finite temperature. Red dots indicate the location of vortex while light- and dark-blue dots indicate the location of antivortices of vorticity $-1$ and $-2$, respectively. Meaning of the arrows and colors as in Fig.~\ref{fig:Figure-5}.}
      \label{fig:Figure-6} 
  \end{figure}
\end{center}

The energetically most favourable configuration presents only two domains [Fig.~\ref{fig:Figure-5}(c)], with larger domain sizes and a smaller number of domain walls. 
However, the energy of this configuration increases with $L$. 
Above $L_{\rm c}$ (long lateral sizes), the two-domain structure [Fig.~\ref{fig:Figure-5}(d)] becomes metastable since the electrostatic energy penalty starts to grow and become dominant. Therefore, we observe new patterns containing two vortices and two antivortices at the interface with SrTiO$_3$, as will be further discussed in Fig.~\ref{fig:Figure-6}. 
Indeed, increasing the temperature we observe how the system is able to escape from this metastable configuration and transit to a state with four domains, as shown in Fig.~\ref{fig:Figure-5}(b). 
The new vortices formed at the interface between PbTiO$_{3}$ and SrTiO$_{3}$ layers, shown in Fig.~\ref{fig:Figure-5}(d), serve as nucleation points of a new down (respectively, up) domains dividing the already existing up (respectively, down) polarization regions. The formation of these new domains reduces the polarization charges generated at the interface.

Assuming in our simulations the in-plane lattice constant of SrTiO$_{3}$~\cite{Shafer-18}, and at low-enough temperatures ($T < $ 50 K), this state is a long-lived metastable phase.  The new topological defects formed at the interface are not able to propagate and close the new domain. Increasing the temperature (beyond 90 K), or inducing compressive strain (beyond -0.5 \%) on the sample, the energy barrier can be overcome and we observe the formation of new domains. 
Interestingly, the transition to the optimal domain configuration is not observed at high domain-densities. Below $L_{\rm c}$ the system is trapped in a polarization wave and cannot transit to a lower-density domain configuration by means of increasing temperature. This asymmetry comes as a consequence of the different nature between the electrostatic and gradient energy penalties. 

Molecular dynamics simulations at constant temperature show how the recombination of vortex-antivortex pairs is the driving mechanism for the domain propagation through the sample until the new domain is completely formed.
In Fig.~\ref{fig:Figure-6}(a) we show in detail the case of a $n=14$, $L=28$ at constant temperature $T=90$ K and slight compressive strain of $-0.5$\%. There we can notice the balance of vortex and antivortex defects resulting in a zero net vorticity on the supercell as stated by the Poincaré-Hopf theorem for our specific periodic boundary conditions. The antivortex textures are mostly formed at the SrTiO$_3$ layers where the magnitude of the polarization and the concomitant electrostatic energy of head-to-head and tail-to-tail domains is smaller. This is in accordance with first-principles calculations \cite{Aguado-Puente-12}. 

In Fig.~\ref{fig:Figure-6}(b) we analyze the time evolution of a portion within the up domain of the same superlattice [see dashed square in Fig.~\ref{fig:Figure-6}(a)], in a region where new polarization vortices have been formed at the interfaces between the PbTiO$_{3}$ and the SrTiO$_{3}$ layers. 
The presence of two vortices (red circles) and an antivortex (light-blue circle) is clearly observed both at the top and the bottom interface. 
Starting from this configuration these vortices and antivortices change their shapes in order to reduce the total energy of the system. 
First, the vortices elongate, while keeping their centers essentially at the same positions. 
This comes with two main consequences. 
First, locally, the number of unit cells with down polarization increases with time [from three at the initial configuration to four at 300 or 400 fs, or even five at 550 fs, see dashed ovals at Fig.~\ref{fig:Figure-6}(b)]. Second, the region where the local polarization points in-plane to close the vortices moves towards the center of the PbTiO$_{3}$ layer, and so does the center of the antivortices. 
At 550 fs, the two antivortices merge to form an antivortex with topological charge -2 at the center of the PbTiO$_{3}$ layer [dark-blue point in Fig.~\ref{fig:Figure-6}(b)]. The field disturbance doubles its charge with a high energetic cost [in accordance with Kosterlitz-Thouless analysis in the sample XY model~\cite{Kosterlitz73,Kosterlitz74} where the energy of the vortices increases (quadratically) with the vorticity].
This is the reason why this state is very short-lived in the molecular dynamic simulations.
Only 50 fs later, it annihilates with two vortices. In this process a new domain with down polarization is formed, together with two new elongated clockwise/counterclockwise pair. 
Finally, the new domain widens till recovering the optimal lateral size determined by the Kittel law. 
\section{Conclusions}
\label{sec:conclusions}
In summary, we theoretically extend the application of the Kittel law to the polar vortex phase in (PbTiO$_3$)$_n$/(SrTiO$_3$)$_n$ superlattices. For the explored layer thicknesses, the square-root dependence of the domain period with the thickness of PbTiO$_3$ is restored by two different procedures: ($i$) full minimization of the energy where all possible interactions are considered, and  ($ii$) analyzing the harmonic force-constant bands. We find that the harmonic approach predicts narrower domains, which is consistent with the fact that anharmonic effects, like the development of an offset, tend to reduce the depolarizing fields on the structure.

Moreover, studying the phase competition under suboptimal domain widths we showed how at low-domain density new domains can be formed to relax electrostatic constraint. These domains nucleate as vortex/antivortex pair defects at the interfaces with SrTiO$_3$ and propagate through the lattice by means of recombination until the new domains are completely formed. This recombination of vortex/antivortex is driven by the high energy costs of polarization patterns containing vortex/antivortex pairs. 
\acknowledgments
F.G.-O., P.G.-F., and J.J. acknowledge financial support from Grant No. PGC2018-096955-B-C41 funded by MCIN/AEI/10.13039/501100011033 and by ERDF ``A way of making Europe'' by the European Union.
F.G.-O acknowledge financial support from grant FPU18/04661 funded by MCIN/AEI/ 10.13039/501100011033.
J.M.L. was supported by Grant No. PID2021-125543NB-I00 funded by MCIN/ AEI/10.13039/501100011033/ and by ERDF ``A way of making Europe'' by the European Union.
H. A. and J. I. were funded by the Luxembourg National Research Fund through Grant C21/MS/15799044/FERRODYNAMICS.
The authors thankfully acknowledge computing time at Altamira supercomputer and the technical support provided by the Instituto de F\'isica de Cantabria  (IFCA) and Universidad de Cantabria (UC).
The authors would also like to thank Jose \'Angel Herrero for his valuable assistance with the supercomputing environment HPC/HTC cluster “Calderon”, supported by datacenter 3Mares, from Universidad de Cantabria.
\bibliography{kittel}
\end{document}